# A rotational ellipsoid model for solid Earth tide with high precision

**Yongfeng Yang[1]*, Yunfei Zhang[2], Qiang Liu[2,3], Xianqing Lv[2]*, Pu Huang[4]**

[1]Water Resources Comprehensive Development Center, Bureau of Water Resources of Shandong Province, Jinan, 250013, China

[2]Frontier Science Center for Deep Ocean Multispheres and Earth System (FDOMES), Physical Oceanography Laboratory, Ocean University of China, Qingdao 266100, China

[3] College of Engineering, Ocean University of China, Qingdao 266100, China

[4] National Laboratory of Solid State Microstructures and Department of Physics, Nanjing University, Nanjing 210093, China

*Corresponding author: Yongfeng Yang (roufeng_yang@outlook.com);

Xianqing Lv (xqinglv@ouc.edu.cn)

## Abstract

Solid Earth tide represents the response of solid Earth to the lunar (solar) gravitational force. The yielding solid Earth due to the force has been thought to be a prolate ellipsoid since the time of Lord Kelvin, yet the ellipsoid's geometry such as major semi-axis's length, minor semi-axis's length, and flattening remains unresolved. Additionally, the tidal displacement of reference point is conventionally resolved through a combination of expanded potential equations and given Earth model. Here we present a geometric model in which both the ellipsoid's geometry and the tidal displacement of reference point can be resolved through a rotating ellipse with respect to the Moon (Sun). We test the geometric model using 23-year gravity data from 22 superconducting gravimeter (SG) stations and compare it with the current model recommended by the IERS (International Earth Rotation System) conventions (2010), the average Root Mean Square (*RMS*) deviation of the gravity change yielded by the geometric model against observation is 6.47 μGal (equivalent to 2.07 cm), while that yielded by the current model is 30.77 μGal (equivalent to 9.85 cm). The geometric model will greatly contribute to many application fields such as geodesy, geophysics, astronomy, and oceanography.

## 1 Introduction

Tides are typical manifestations of the Earth's systems in response to the gravitational force of the Moon and Sun[1,2]. The response of solid Earth to the force is called body tide or solid Earth tide. Not like ocean tide that is rather spectacular, solid Earth tide is more difficult to observe due to the simultaneous movement of the reference frame and the observer[3,4]. Even so, the tidal signal of solid Earth can be captured with instruments such as horizontal pendulum, tide gauge, gravimeter, Very Long Baseline Interferometry, Global Positioning System[3,5-9], diamagnetic-levitated micro-oscillator[10], and so on.

A theoretical treatment of solid Earth tide commonly incorporates the expanded potential equations and given Earth model to resolve the dimensionless Love-Shida numbers[1,4,11-12]. The Love-Shida numbers represent the ratios between the observation under the forcing and the expectation in the equilibrium tide theory. The existing solid Earth tide theory has been developed into a model highly recommended by the IERS[13-16]. The tidal displacement computed by the current model is being widely utilized for the correction on the deflection of the vertical, satellite orbit, the Interferometric Synthetic Aperture Radar image, tectonic velocity, glacial isostatic adjustment, Global Seismographic Network, and vertical land motion at tide gauges[17-23].The tidal displacement computed by the model is also subtracted from the measured sea surface height when ocean tide models are created[24]. Solid Earth tide may help reveal the Earth's interior structure[25], the Earth's response to ocean tide loading[21], and the correlation between Earthquake and tidal stress[26]. Solid Earth tide computed has been the key for understanding sea level change and variability[27,28], and has been fundamental for establishing the International Celestial Reference Frame[29], the International Terrestrial Reference Frame[30], and for resolving geopotential[31].

Since the time of Lord Kelvin, it has been believed that the yielding solid Earth due to the lunar (solar) gravitational force is a prolate ellipsoid, in which solid Earth is slightly elongated along the Earth-Moon (Sun) line and shortened around the parts remote from the elongation[32]. Nevertheless, the ellipsoid's geometry such as major semi-axis's length, minor semi-axis's length, and flattening are not resolved. The existing solid Earth tide theory has been tested. Most of these studies[33-35] making comparison of theory and observation are limited to the gravimetric factor of specific tidal constituents (e.g., O1, M2, P1, and K1). A tidal constituent is a mathematical expression for the tidal force, and O1, P1 and K1 are the dominant diurnal constituents, while M2 is one of the semidiurnal constituents. The gravimetric factor is a ratio of the observation and expectation and itself is a linear combination of the Love-Shida numbers[9]. Different from this routine, Goodkind[36] constructed two theoretical tide time series plus the local barometric pressure time series to fit the gravimeter signal in order to determine three scalar parameters. Practically, the tidal displacement of a site is treated as a sum of the displacements of all tidal constituents, and the displacement of each constituent is computed through a cosine function that is incorporated by the amplitude factor, the amplitude, and the phase lag. The number of the tidal constituents selected for tidal displacement prediction often reaches tens or hundreds or more. From this side, a comparison of the gravimetric factor does not relate to the tidal displacement, and consequently, the accuracy of tidal displacement predicted by the existing theory remains unclear.

Superconducting gravimeter (SG) data provide the most accurate information about solid Earth tide[17,37]. A detailed description of SGs and their measurements may refer to this book[38]. More than 40 SG stations are presently included in the network of International Geodynamics and Earth Tide Service (IGETS), and their time series cover more than 25 years. The IGETS gravity data are globally available through open access[37,39], and some of them have been used for creating local tide models using harmonic analysis[40-42], and their adaptability was well discussed[43]. The objects of this study are twofold, firstly, to present a geometric model with hope of improving our understanding of solid Earth tide, and secondly, to examine the accuracy of the current model recommended by the IERS conventions (2010).

## 2 Method and Result

The Earth can be considered as a solid sphere with a thicker core, encompassed by a thinner layer of water and atmosphere[44]. This approximation allows to neglect any material variations within the Earth and any discrepancies in radius, assuming it to be a symmetrical, rotating, elastic, homogeneous, and oceanless sphere. The Earth participates in two curved motions in the solar system: one is the Earth orbiting around the barycenter of the Earth-Moon system and the other is the Earth-Moon system orbiting around the Sun (Fig. 1(a)). A fuller description of the motions of the Earth, Moon, and Sun can be found in these works[45-46]. Mechanically, as illustrated in Figure 1 (*b*), a curved motion in space can generate a centrifugal force $f$ for a sphere. The centrifugal force is counterbalanced by a centripetal force $F$, which is a pull force exerted along the rod. The sphere's response to a combination of these opposing forces elongates its body along the line of force. Similarly, the two curved motions generate two centrifugal forces, $f_1$ and $f_2$, for solid Earth. These two centrifugal forces $f_1$ and $f_2$ are counterbalanced by the gravitational force $F_1$ from the Moon and gravitational force $F_2$ from the Sun, respectively. It is these force balances that enable the Earth to steadily move around the barycenter of the Earth-Moon system and around the Sun. The Earth's response to a combination of these opposing forces $f_1$ ($f_2$) and $F_1$ ($F_2$) slightly elongates the Earth's body along the Earth-Moon (Sun) line and shortens the body midway (Fig. 1(*c* and *d*)). We assume that the deformation can be geometrically represented with a prolate ellipsoid whose major axis points to the Earth-Moon (Sun) line and center coincides with the center of the Earth.

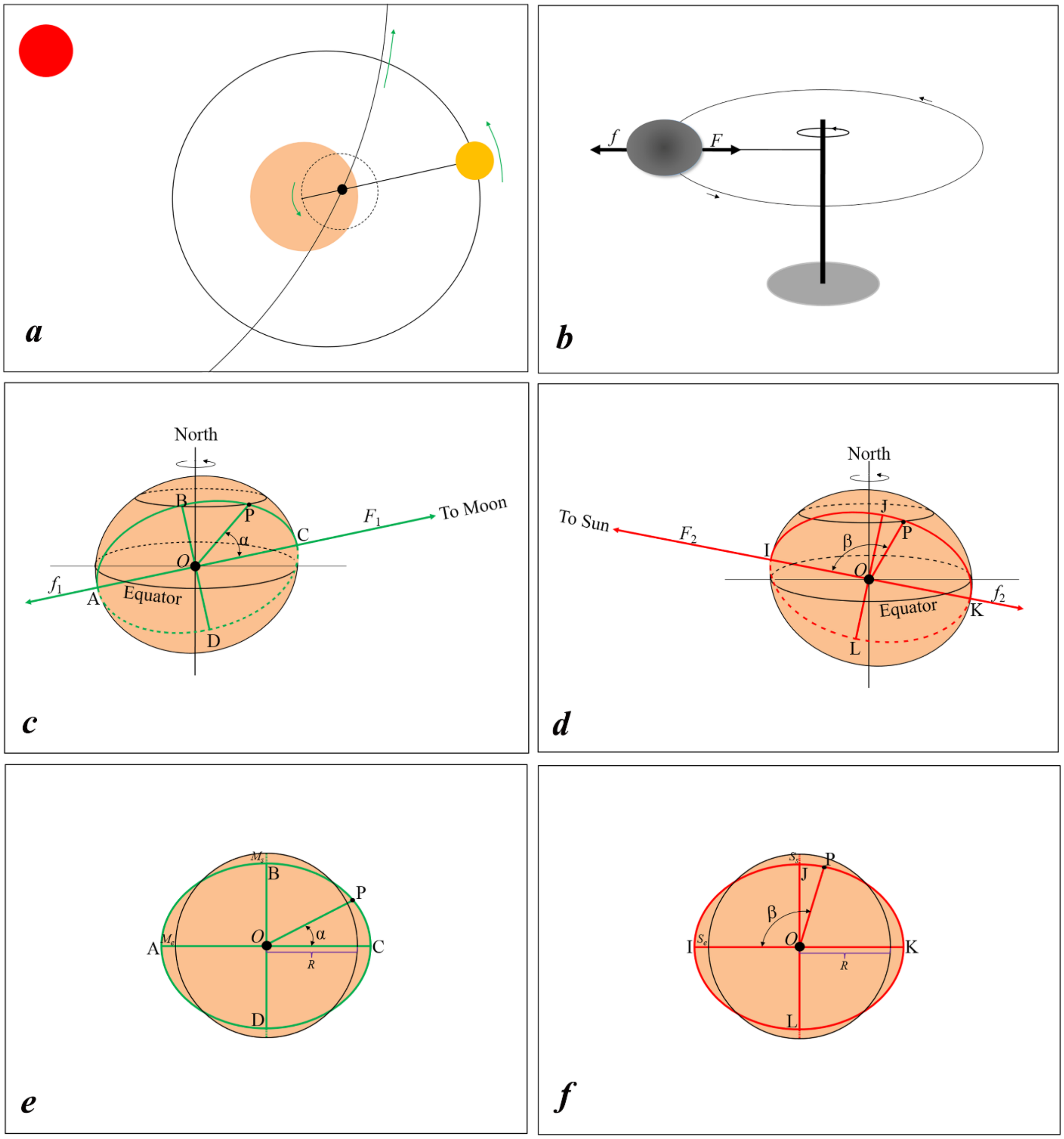

**Figure 1. Combined effect of opposing forces on solid Earth.** (*a*), the Earth's motions around the barycenter of the Earth-Moon system and around the Sun. (*b*), a sphere's deformation due to the combination of pull force and centrifugal force. (*c* and *d*), the resultant prolate ellipsoid due to a combination of the two opposing forces, i.e., centrifugal forces $f_1$ ($f_2$) and gravitational force $F_1$ ($F_2$). α and β are the lunar and solar angles of site P relative to the Earth's center *O*, respectively. (*e* and *f*), the sections dissected from the prolate ellipsoid passing site P and the major axis AC

(IK). Black circle denotes the original shape of section ABCD (IJKL). $M_e$ ($S_e$) and $M_s$ ($S_s$) denote the elongation and shortening in the prolate ellipsoid due to the Moon (Sun), respectively.

As exhibited in Figure 1($c$($d$)), at the moment we use a section that passes a site and the prolate ellipsoid's major axis to cut the ellipsoid's body, the resultant intersection line at the ellipsoid's surface would be an ellipse. As the Earth spins around its axis, the site is entrained to move. The relative positions between the site, the Earth's center, and the Moon are timely varying. After a period of time, we use a section that passes the site and the prolate ellipsoid's major axis to cut the ellipsoid's body again, the resultant intersection line at the ellipsoid's surface would be another ellipse. We find, at any time the resultant intersection line at the ellipsoid's surface due to the cutting would be an ellipse. Consequently, the distance of the site from the Earth's center at a time can be expressed through a geometry of the ellipse (Fig. 1($e$ and $f$)). Taking into account the co-existence of the two prolate ellipsoids due to the Moon and Sun, the combined vertical displacement of site P relative to the Earth's center at a time $t$ can be written as $H_{(t)}=H_{m(t)}+H_{s(t)}$, where

$$\begin{aligned} H_{m(t)} &= \sqrt{(R+M_e)^2\cos^2\alpha + (R-M_s)^2\sin^2\alpha} - R \\ H_{s(t)} &= \sqrt{(R+S_e)^2\cos^2\beta + (R-S_s)^2\sin^2\beta} - R \end{aligned} \tag{1}$$

Where $H_{m(t)}$ ($H_{s(t)}$) is the vertical displacement of a site relative to the Earth's center in the prolate ellipsoid due to the Moon (Sun) at the time. $R$ is the mean radius of solid Earth, $M_e$ ($S_e$) and $M_s$ ($S_s$) are the elongation in the major semi-axis and the shortening in the minor semi-axis due to the Moon (Sun) at the time, respectively. ($R+M_e$) and ($R-M_s$) are the major semi-axis's length and minor semi-axis's length of the ellipsoid due to the Moon at the time, whereas ($R+S_e$) and ($R-S_s$) are the major semi-axis's length and minor semi-axis's length of the ellipsoid due the Sun at the time. $\alpha$ and $\beta$ are the lunar and solar angles of site P relative to the Earth's center at the time, and they can be calculated through a standard formula in spherical trigonometry[47]: $\cos\alpha=\sin\sigma\sin\delta_m+\cos\sigma\cos\delta_m\cos C_{mm}$, $\cos\beta=\sin\sigma\sin\delta_s+\cos\sigma\cos\delta_m\cos C_{ms}$. Where $\sigma$, $\delta_m$, $\delta_s$, $C_{mm}$, and $C_{ms}$ are the geographic latitude of site P, the declination of the Moon, the declination of the Sun, the hour angle of site P with respect to the Moon, and the hour angle of site P with respect to the Sun, respectively.

$M_e$ ($S_e$) and $M_s$ ($S_s$) can be developed into $M_e=K_{me}P_m\cos^2\delta_m$ ($S_e=K_{se}P_s\cos^2\delta_s$) and $M_s=K_{ms}P_m\cos^2\delta_m$ ($S_s=K_{ss}P_s\cos^2\delta_s$), where $P_m=D_{me}^2/D_m^2$ ($P_s=D_{se}^2/D_s^2$) is a distance factor that relates to the Earth and Moon (Sun), $D_{me}$ ($D_{se}$) is the mean distance between the Earth and Moon (Sun), $D_m$ ($D_s$) is the temporary distance between the Earth and Moon (Sun) at the time. $K_{me}$ ($K_{se}$) and $K_{ms}$ ($K_{ss}$) are undetermined parameters of elongation and shortening in the prolate ellipsoid due to the Moon (Sun), $\cos^2\delta_m$ ($\cos^2\delta_s$) denotes a latitude factor that relates to the position of the Moon (Sun).

Equation (1) indicates that, if the vertical displacement of a site is measured, we may use the measured data and ephemeris data as input and take a least-squares fitting to resolve the undetermined parameters $K_{me}$ ($K_{se}$) and $K_{ms}$ ($K_{ss}$). Furthermore, the elongation $M_e$ ($S_e$) and the shortening $M_s$ ($S_s$) in the prolate ellipsoids can be determined with these known parameters and ephemeris data.

The vertical displacement of a site can be measured by GPS whose positioning precision has presently reached a millimeter level with a static solution[38]. But if a dynamic solution is applied, the precision will be only in one centimetre level[48], which is not good for resolving these parameters. Instead, the vertical displacement corresponds to gravity change, and gravity change has been widely measured by SGs whose precision is less than 0.05μGal, and therefore, the gravity change data are ideal for resolving these parameters. The measured gravity change as a whole include the effects caused by solid Earth tide, atmosphere pressure, pole motion, ocean tide loading, and so on[38]. This means that the gravity change caused by solid Earth tide must be firstly separated from the measured gravity change and then used for the solution of our model. Here, we select time series of four SG stations (Canberra, Bad Homburg, Apache Point, and Sutherland) from IGETS[37], and these stations are located in Australia, Europe, America, and Africa, respectively. The gravity data in IGETs are recorded with three levels: the level 1 are raw without any correction, the level 2 are corrected for instrumental perturbations, and the level 3 are those after particular geophysical corrects had been separated into solid Earth tide, atmosphere load, rotation, drift, ocean tide loading, residuals, and so on[37]. This separation allows us to directly utilize the gravity change caused by solid Earth tide. The time covering of the selected time series is from August 1, 2014, to December 30, 2014. The latitude and longitude of these stations are listed in Table 1. As the unit of gravity change differs from that of vertical displacement, a transformation between two kinds of data was considered. The transformation from vertical displacement to gravity change can be approximately expressed as

$$\Delta g = (1/(H+R)^2 - 1/R^2)GM_e \quad (2)$$

where $H$ and $\Delta g$ are vertical displacement and its resultant gravity change, respectively. $G$, $M_e$, $R$ are the gravitational constant, mass of the Earth, and the mean distance of a SG station from the Earth's center, respectively. We here assume that the mean distance is equal to the Earth's mean radius. $G = 6.67\times10^{-11} m^1 s^{-2}$, $M_e = 5.97\times10^{24}$ kg[49], and $R$ = 6371 km[50].

In the equation (2), we use the Earth's mean radius to replace the mean distance of a SG station from the Earth's center, but the mean distance is not equal to the Earth's mean radius, this replacement deserves a discussion. When $R$ = 6356 km (if the SG station's mean distance is equal to the pole radius) or $R$ = 6378 km (if the SG station's mean distance is equal to the equator radius), for a vertical displacement $H$=0.6 m, the equation (2) can result in a gravity change error of -1.31 μGal or 0.61 μGal relative to the vertical displacement computed with the Earth's mean distance $R$ = 6371 km. Most of the SG stations in the network of IGETS have gravity change of greater than 100 μGal, the gravity change error computed through the equation above is on one microgal level and therefore can be neglected.

Ephemeris data, including the declination of the Moon (Sun) and the average and temporary distance of the Moon (Sun) from Earth, were obtained from NASA JPL horizons system. The fitted results for $K_{\mathrm{me}}$, $K_{\mathrm{ms}}$, $K_{se}$, and $K_{ss}$ are 41.33 cm, 21.41 cm, 15.93 cm, and 13.03 cm, respectively. Taking into account the varying inclination of the Moon (Sun) and the temporary distance between the Earth and Moon (Sun), we estimate that during a period of time (from 1990 to 2030), the elongation $M_{\mathrm{e}}$ ($S_{\mathrm{e}}$) and shortening $M_{\mathrm{s}}$ ($S_{\mathrm{s}}$) of solid Earth due to the Moon (Sun) range from 49.75 cm to 29.77 cm (16.09 cm to 12.98 cm) and from 25.77 cm to 15.42 cm (13.16 cm to 10.62 cm), respectively. The variations of $M_{\mathrm{e}}$, $M_{\mathrm{s}}$, $S_{\mathrm{e}}$, and $S_{\mathrm{s}}$ are typically compared in Fig. 2. Given the Earth's mean radius $R$=6371000.00 m, the major semi-axis's length and minor semi-axis's length in the ellipsoid due to the Moon vary from 6371000.50 m to 6371000.30 m and from 6370999.85 m to 6370999.74 m, and the major semi-axis's length and minor semi-axis's length in the ellipsoid due to the Sun vary from 6371000.16 m to 6371000.13 m and from 6370999.89 m to 6370999.87 m. The flattening of an ellipsoid is written as $f=(a-c)/a$, where $f$, $a$, and $c$ are the flattening of the ellipsoid, the major semi-axis's length, and the minor semi-axis's length, respectively. As a result, the flattening for the ellipsoid due to the Moon (Sun) ranges from $7.09*10^{-8}$ ($3.71*10^{-8}$) to $11.85*10^{-8}$ ($4.59*10^{-8}$).

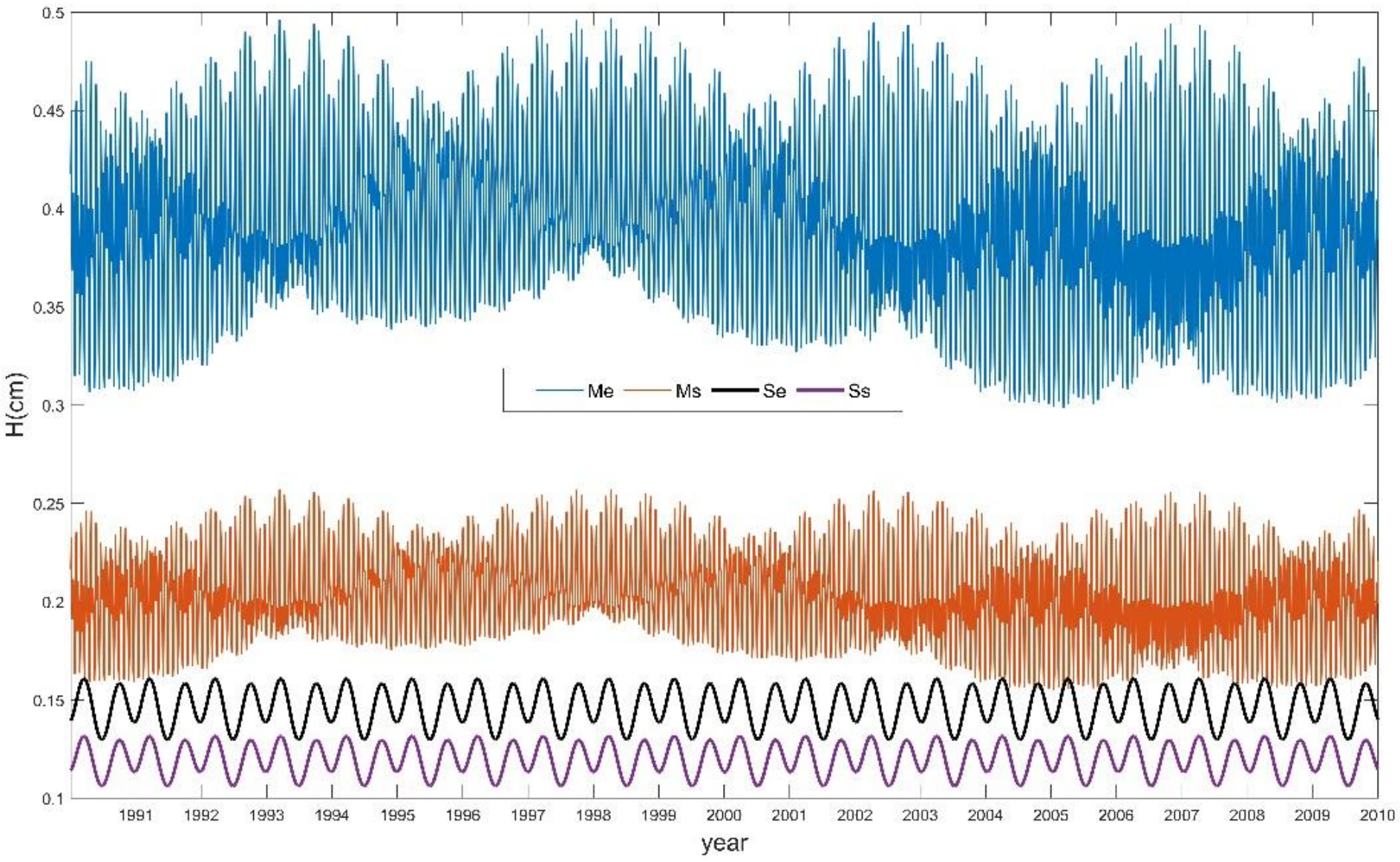

**Figure 2. Variations of the elongation $M_e$ ($S_e$) and shortening $M_s$ ($S_s$) of solid Earth from 1990 to 2010.** $M_e$ ($S_e$) and $M_s$ ($S_s$) are calculated through $M_e = K_{me} P_m \cos^2\delta_m$ ($S_e = K_{se} P_s \cos^2\delta_s$) and $M_s = K_{ms} P_m \cos^2\delta_m$ ($S_s = K_{ss} P_s \cos^2\delta_s$), and $P_m$ and $P_s$ are the distance factors that relates to the Earth and Moon (Sun), $D_{me}$ ($D_{se}$) are the mean distance between the Earth and Moon (Sun), $D_m$($D_s$) is the temporary distance between the Earth and Moon (Sun).

Taking the equation (1) and ephemeris data, one may compute the tidal displacement of any site on the Earth's surface at any time. Gravity data of 22 SG stations from IGETS are used to test the geometric model, and a geographic distribution of these 22 SG stations is shown in Figure 3. Similarly, the gravity change caused by solid Earth tide are selected from the level 3 of the IGETs. These stations are located around the world, ensuring their representativeness within the network. The selected time series of each of these stations have a continuous coverage of 6 months, and the total covering of the selected data spans from January 1998 to June 2021. For each station, the selected data are hourly, totally 4320 data points are include for analysis. To examine the strength of the geometric model, we compare it with the current solid Earth tide model recommended by the IERS conventions (2010). The current model has been developed/incorporated into some softwares such as Solid[51], pyTMD[3], and Etideload[52]. Solid is an implementation of the solid earth tide computation documented in the IERS Conventions (2003). The solid Earth tide computation incorporated in pyTMD is based on the IERS conventions (2010). Since the solid Earth tide computation documented in the IERS conventions (2003) is identical to that documented in the IERS conventions (2010), the solid Earth tide computed by Solid is equivalent to that computed by pyTMD. The solid Earth tide computation incorporated in Etideload is mostly based on the IERS conventions (2010), but the effect of planets on Earth tide is included. Hence, we select Solid and pyTMD as representative of the current model. The solid Earth tide computed by the current model is the displacement, too. To ease the comparison, we use the equation (2) to transform the vertical displacement computed by the geometric model and the current model into gravity change. The RMS deviation for a SG station can be written as follows

$$RMS = \sqrt{\sum_i^m (g_m(i) - g_o(i))^2 / m} \tag{3}$$

where $g_{\mathrm{m}}(i)$ represents the gravity change computed by the geometric model or the current model (Solid (pyTMD) and Etideload), $g_{\mathrm{o}}(i)$ represents the observed gravity change at the SG station, and $m$ is the total number of observations used in the analysis. The detailed information of the selected time series of these stations and the *RMS* deviation of various models are listed in Table 1.

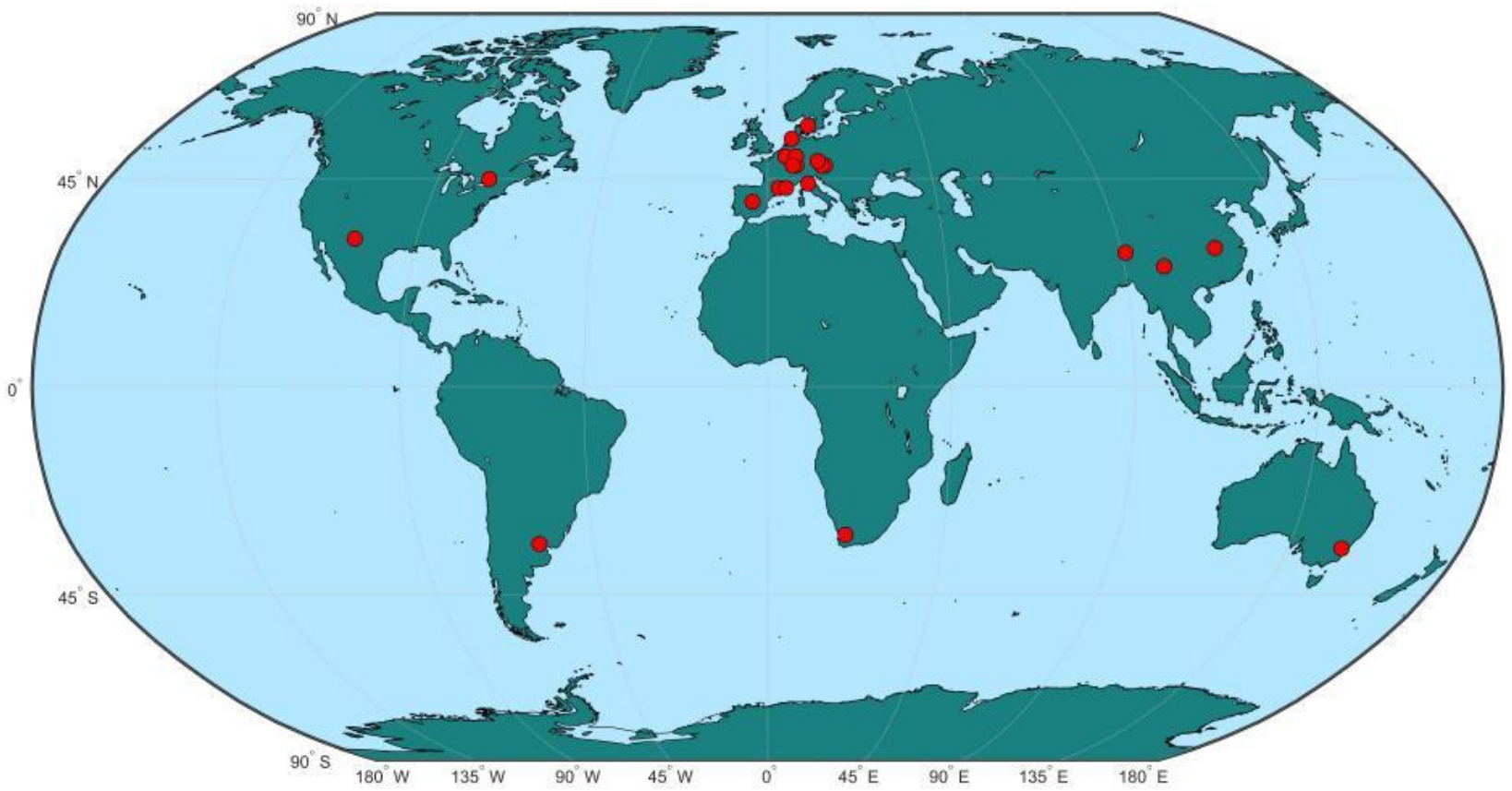

**Figure 3. A geographic distribution of the 22 SG stations.**

**Table 1. Time series of 22 SG stations and the *RMS* deviation of various models against observation**

| SG Station | Latitude | Longitude | Time Span | *RMS* (μGal) | | |
|---|---|---|---|---|---|---|
| | | | | Etideload | Solid (pyTMD) | Geometric Model |
| Vienna | 48.25° | 16.36° | 19980101-19980630 | 27.82 | 27.97 | 4.92 |
| Wuhan | 30.52° | 114.49° | 19980101-19980630 | 31 | 31.48 | 5.53 |
| Apache Point | 32.78° | 254.18° | 20160101-20160630 | 29.98 | 30.55 | 5.3 |
| Conrad | 47.93° | 15.86° | 20160101-20160630 | 27.09 | 27.68 | 4.84 |
| Lhasa | 29.65° | 91.04° | 20160101-20160630 | 30.79 | 31.26 | 5.2 |
| Lijiang | 26.90° | 100.23° | 20160101-20160630 | 31.96 | 32.35 | 5.5 |
| La Plata | -34.87° | 301.86° | 20180101-20180630 | 27.21 | 27.77 | 5.02 |
| Medicina | 44.52° | 11.65° | 20180101-20180630 | 30.94 | 31.56 | 6.4 |
| Rochefort | 50.16° | 5.23° | 20180101-20180630 | 28.6 | 29.19 | 5.65 |
| Schiltach | 48.33° | 8.33° | 20180101-20180630 | 28.51 | 29.05 | 5.52 |
| Aubure | 48.22° | 7.20° | 20210101-20210630 | 28.46 | 29.03 | 5.54 |
| Bad Homburg | 50.23° | 8.61° | 20210101-20210630 | 30.48 | 31.49 | 7.34 |
| Canberra | -35.32° | 149.01° | 20210101-20210630 | 31.69 | 32.55 | 8.65 |
| Cantley | 45.59° | 284.19° | 20210101-20210630 | 30.95 | 32.07 | 8.08 |
| Helgoland | 54.18° | 7.89° | 20210101-20210630 | 29.98 | 30.83 | 6.18 |
| Larzac | 43.97° | 3.22° | 20210101-20210630 | 30.33 | 31.38 | 7.63 |
| Onsala | 57.39° | 11.93° | 20210101-20210630 | 31.61 | 32.32 | 6.25 |
| Pecny | 49.91° | 14.79° | 20210101-20210630 | 30.57 | 31.54 | 7.14 |
| Rustrel | 43.94° | 5.48° | 20210101-20210630 | 30.44 | 31.49 | 7.66 |
| Strasbourg | 48.62° | 7.68° | 20210101-20210630 | 30.45 | 31.45 | 7.26 |
| Sutherland | -32.38° | 20.81° | 20210101-20210630 | 32.03 | 32.88 | 9.09 |
| Yebes | 40.52° | 356.91° | 20210101-20210630 | 30.05 | 31.11 | 7.72 |
| | | | Mean | 30.04 | 30.77 | 6.47 |

We find that the geometric model is significantly superior to the existing models, the average RMS deviation of the gravity change computed by the geometric model against observation across the 22 SG stations is 6.47 μGal, the average RMS of the gravity change computed by Solid (pyTMD) is 30.77 μGal, while the average RMS of the gravity change computed by Etideload is 30.04 μGal. One microgal of gravity change is equivalent to about 3.2 mm of relative height change[53], these RMS deviations of gravity change thereby correspond to height change of 2.07 cm, 9.85 cm, and 9.61 cm, respectively. The comparison is typically displayed with four SG stations (Vienna, Aubure, Apache Point, and La Plata) (Fig.4), showing that the gravity change computed by the geometric model is more consistent with observed data than the existing models. The gravity change computed by the current model is overwhelmingly less than the observed gravity change, and at the time of high tide, the difference between the two may rise up to 70 μGal (equivalent to about 22 cm).

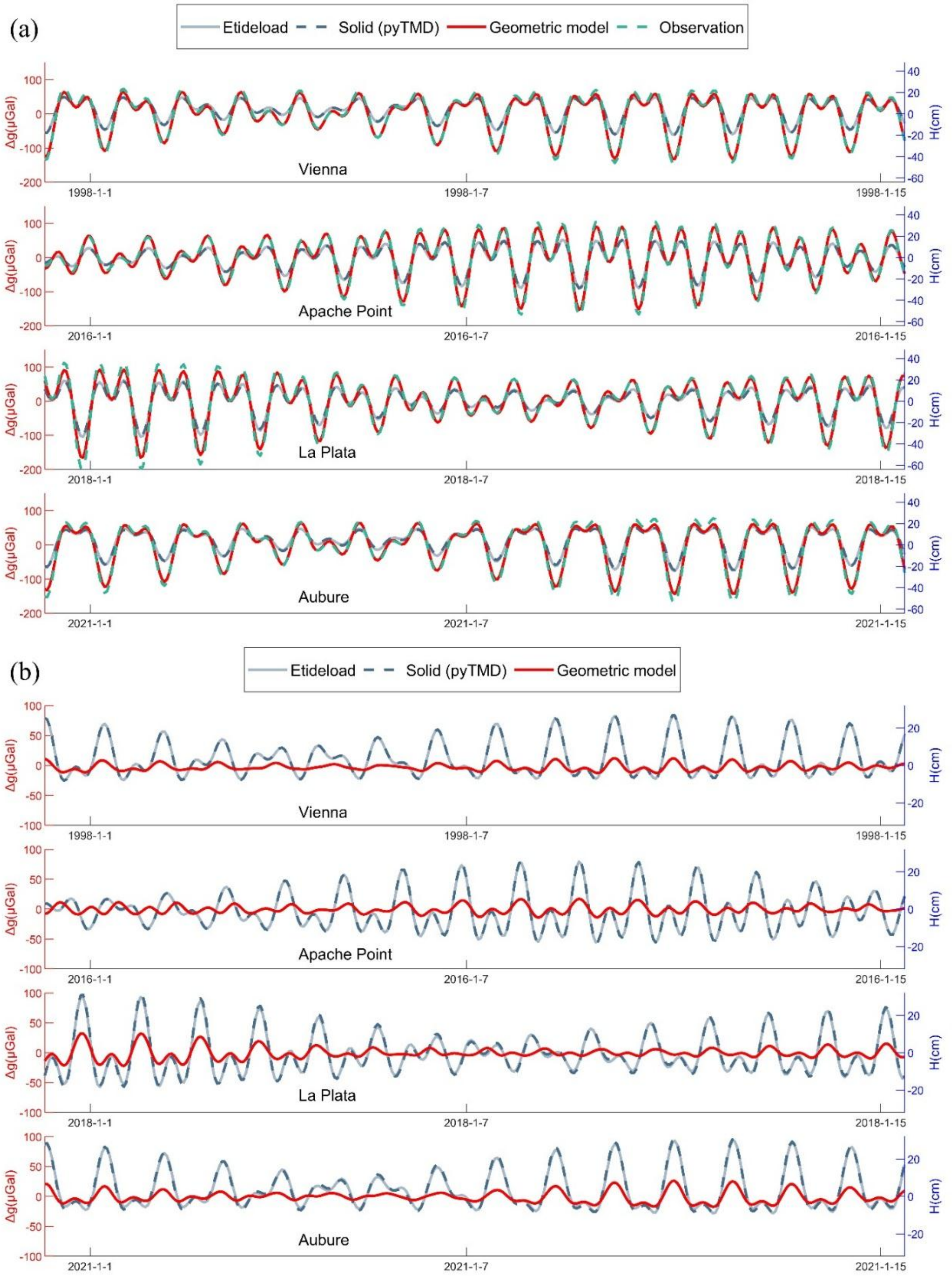

**Figure 4. Gravity change comparison between various models.** (a) Gravity change computed by the models at four SG stations. (b) Gravity change difference between the models and observation.

We show that from 1990 to 2030, the averaged $M_e$ is about 2.7 times of the average $S_e$, and the averaged $M_s$ is about 1.7 times of the average $S_s$. These ratios determine solid Earth to be dominantly elongated along the Earth-Moon line and shortened around the parts remote from the elongation. This geometry of deformation implies that, if a site enters the elongation, it will experience a fall in gravity, and if the site enters the shortening, it will experience an increase in gravity. To examine this expectation, we construct a correlation between lunar angle and gravity variation for 20 SG stations (Fig. 5). Results uniformly indicate a decline in gravity during lunar angles of 0°-45° and 135°- 180°, while an increase in gravity during lunar phases of 45°-135°, supporting the existence of the elongated solid Earth.

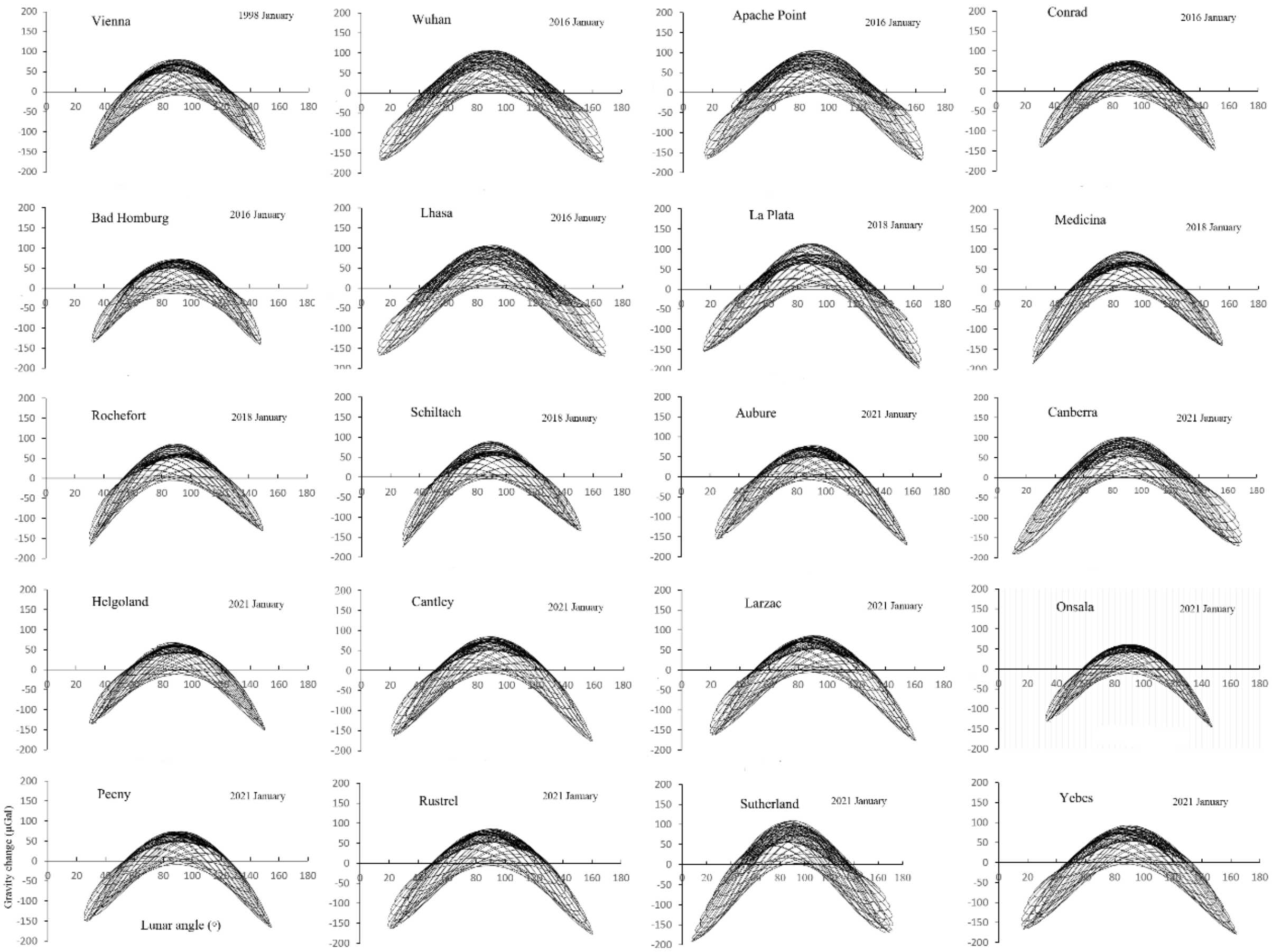

**Figure 5. Gravity change out to lunar angle.** Gravity data of 20 SG stations are selected from IGETs[37]. Lunar angle is the angle of the SG station and the Moon relative to the Earth's center.

## 3 Discussion

Essentially, the current model uses the tidal force (potential) and the response of solid Earth to it (i.e., the Love-Shida numbers) to compute the tidal displacement. Due to the spatial variation of the tidal force, and due to the complexity of the Earth's structure and the materials within it, a precise understanding of the response is hard to reach. Moreover, the Love-Shida numbers are only constrained by limited observations, this cannot guarantee their adaptability for a global scale. Hence, the current model would be difficult to realize a high accuracy. This issue may find reference from astronomy. The Copernicus's heliocentric model presented that the planets are revolving around the Sun, not the other way around. However, the details of these movements, such as their distance from the Sun and orbital speed, are computed through the Kepler's elliptical model of planetary orbit. Although the planet's elliptical orbit is widely known to be a consequence of the Sun's gravitation force, it is difficult to use the force to directly compute the movement of planet around the Sun.

The low accuracy of the current model has imposed challenge on its application. It is reported that the best ocean tide models available today have the accuracy of a few centimeters in open ocean[54]. In creating these models, the solid Earth tide computed by the current model needs to be subtracted from the measured sea surface height so as to obtain the measured ocean tide. The measured sea surface height is expressed with $h_{ssh}=h_{MSS}+h_s+e$, where $h_{MSS}$ is the mean sea surface, $h_s$ is the (instantaneous) sea surface height anomaly that is corrected by the potential satellite bias, the instrument correction, the solid Earth tide, the ocean tide, and so on, and $e$ is the error in these corrections and the (unmodeled) general ocean circulation signal[24]. Hence, the observed ocean tide obtained through the expression is overestimated by 9 cm due to the inadequacy of the current model. Further, the observed ocean tide is used to resolve the parameters of tidal constituents, with these known parameters, the ocean tide model is eventually built and used to make prediction. Our demonstration of ocean tide model reveals that the accuracy of a few centimeters for the best ocean tide models cannot be realistic. NOAA Climate.gov recently reported that the rate of global sea level rise has increased from 1.4 mm per year throughout most of the twentieth century to 3.6 mm per year from 2006–2015. The sea level rise is determined through tide gauge and satellites that relate to vertical land motion, and the vertical land motion is corrected by the solid Earth tide computed by the current model. For accurate determination of sea level change, vertical land motion is required to have standard errors that are 1 order of magnitude lower than the contemporary climate signals of 1 to 3 mm/yr observed on average in sea level records from tide gauges or satellites[27]. Inadequate correction of solid Earth tide on vertical land motion may flood a vertical land motion of a few millimeters per year. It is no doubt, the low accuracy of the current model may discount any of precise ground- and spatial-based measurements.

## 4 Conclusions

The yielding Earth due to the lunar (solar) gravitational force has long been thought to be a prolate ellipsoid, in which solid Earth is slightly elongated along the Earth-Moon (Sun) line and shortened around the parts remote from the elongation. Yet, the ellipsoid's geometry such as major semi-axis's length, minor semi-axis's length, and flattening remains unresolved. Additionally, the tidal displacement of solid Earth is conventionally resolved through a combination of expanded potential equations and given Earth model. In this study, we represent a geometric model in which both the geometry of ellipsoid and the tidal displacement of solid Earth can be resolved through a rotating ellipse with respect to the Moon (Sun). We find that during a period of time (from 1990 to 2030), the ellipsoid's flattening due to the lunar (solar) gravitational force varies from $7.09*10^{-8}$ ($3.71*10^{-8}$) to $11.85*10^{-8}$ ($4.59*10^{-8}$). We test the geometric model using 23-year gravity data from 22 superconducting gravimeter (SG) stations and compare it with the current model recommended by the IERS (International Earth Rotation System) conventions (2010), the average Root Mean Square (*RMS*) deviation of the gravity change yielded by the geometric model against observation is 6.47 μGal (equivalent to 2.07 cm), while that yielded by the current model is 30.77 μGal (equivalent to 9.85 cm). The geometric model represents a significant advance in understanding and predicting solid Earth tide, and will greatly contribute to many application fields such as geodesy, geophysics, astronomy, and oceanography.

**Acknowledgments:** Authors declare that they have no competing interests. This research was supported by the National Natural Science Foundation of China (Grant No. 42076011) and the National Key Research and Development Program of China (Grant No.2022YFC3105003). We thank Tyler C. Sutterley, Mao Zhou, and Chuanyin Zhang for help in producing solid earth tide data through different softwares, and Jean-Paul Boy for help in obtaining superconducting gravimeter data from IGETS. The authors are very grateful to Jon Giorgini for help in extracting ephemeris data from NASA JPL horizons system. The authors are also grateful to Dennis Milbert for his clarification of the IERS conventions and Solid earth tide software, and to Heping Sun, Gerhard Jentzsch, and Christian Voigt for their valuable comments.

**Author Contributions:** YFY conceived the model and drafted the manuscript. YFZ and YFY contributed to the parameter fitting and validation of the model. QL and PH carried out the logical and expressive modifications of the paper and actively engaged in discussions throughout the progress of the paper. XQL was responsible for the overall planning of the research and improved the model. All authors participated in the revision of the paper.

**Data Availability Statement:** The superconducting gravity data used in this research are available through Voigt et al., 2016 (ftp://igetsftp.gfz-potsdam.de). The ephemeris data used are obtained from NASA JPL horizons system (https://ssd.jpl.nasa.gov/horizons/).